\documentclass[useAMS,usenatbib]{mn2e}
\usepackage{epsfig,graphicx,latexsym,amsmath,amssymb}
\usepackage{natbib}
\usepackage{hyperref}
\usepackage{mathrsfs}
\usepackage{rotating,url}

\bibpunct[,]{(}{)}{;}{a}{}{,}

\newcommand {\Msun}{\ensuremath{M_{\odot}}}

\newcommand {\kms}{\ensuremath{\mathrm{km\,s}^{-1}}}

\newcommand{\rem}[1]{} 
\newcommand{\gta}{\mathrel{\raise.3ex\hbox{$>$}\mkern-14mu
             \lower0.6ex\hbox{$\sim$}}}
\newcommand{\lta}{\mathrel{\raise.3ex\hbox{$<$}\mkern-14mu
             \lower0.6ex\hbox{$\sim$}}}

\title[Tidal disruptions of separated binaries]
      {Tidal disruptions of separated binaries in galactic nuclei}

\author[P. Amaro-Seoane, M. Coleman Miller and Gareth F. Kennedy
        ]
{Pau Amaro-Seoane$^{1,\,2}$
\thanks{E-mail: Pau.Amaro-Seoane@aei.mpg.de (corresponding author)}, 
M. Coleman Miller$^{3}$ \& Gareth F. Kennedy$^{4}$
   \\
$^{1}$Max Planck Institut f\"ur Gravitationsphysik
(Albert-Einstein-Institut), D-14476 Potsdam, Germany
\\
$^{2}$Institut de Ci{\`e}ncies de l'Espai (CSIC-IEEC), Campus UAB,
Torre C-5, parells, $2^{\rm na}$ planta, ES-08193, Bellaterra,
Barcelona, Spain
\\
$^{3}$University of Maryland Department of Astronomy and
Joint Space-Science Institute, College Park, MD, 20742-2421, USA\\
$^{4}$Institut de Ci{\`e}ncies del Cosmos, Facultat de F{\'\i}sica
Marti i Franqu{\`e}s, 1
E-08028 Barcelona, Spain
}

\begin{document}

\date{draft \today}

\pagerange{\pageref{firstpage}--\pageref{lastpage}} \pubyear{2010}

\maketitle

\label{firstpage}

\begin{abstract}
Several galaxies have exhibited X-ray flares that are consistent
with the tidal disruption of a star by a central supermassive black
hole.  In theoretical treatments of this process it is usually
assumed that the star was initially on a nearly parabolic orbit
relative to the black hole.  Such an assumption leads in the
simplest approximation to a $t^{-5/3}$ decay of the bolometric
luminosity and this is indeed consistent with the relatively poorly
sampled light curves of such flares.  We point out that there is
another regime in which the decay would be different: if a binary is
tidally separated and the star that remains close to the hole is
eventually tidally disrupted from a moderate eccentricity orbit, the
decay is slower, typically $\sim t^{-1.2}$.  As a result, careful
sampling of the light curves of such flares could distinguish
between these processes and yield insight into the dynamics of
binaries as well as single stars in galactic centres.  We explore
this process using three-body simulations and analytic treatments
and discuss the consequences for present-day X-ray detections and
future gravitational wave observations.
\end{abstract}


\section{Introduction}
\label{sec_intro}

In the past few years, several galaxies have exhibited X-ray/UV flares
consistent with the tidal disruption of a star by a supermassive black hole
(SMBH; for flare observations see \citealt{DonleyEtAl02,DogielEtAl09,GezariEtAl09}).
These candidate disruptions are relevant to the fueling of some active galactic
nuclei (particularly low-mass ones; see \citealt{WangMerritt04}) and contain
important information about stellar dynamics in the centers of galaxies.  In
addition, they are related to one of the processes believed to lead to extreme
mass ratio inspirals (EMRIs), in which a stellar-mass object spirals into a
supermassive black hole; EMRIs are thought to be among the most promising
sources for milliHertz gravitational wave detectors such as the {\it Evolved Laser
Interferometer Space Antenna} \citep[LISA, see][]{Amaro-SeoaneEtAl07,Amaro-SeoaneEtAl2012}.

Analyses of stellar tidal disruptions have focused on stars whose orbits are
nearly parabolic relative to the SMBH \citep{Rees88}.  In this case, roughly
half the stellar material becomes unbound and the rest rains down on the SMBH
with a rate that, for simplified stellar structure, scales with the time $t$
since disruption as ${\dot M}\sim t^{-5/3}$ (this is expected at late times
even for more realistic structure; see \citealt{LodatoEtAl09}).  

There is, however, another possible path to disruptions.  Binaries that
get close enough to a SMBH can be tidally separated without destroying
either star. The result is that one star becomes relatively tightly bound
to the SMBH whereas the other is flung out at high speed.  The bound star
will undergo dynamical interactions and its orbit will also shrink and
circularise due to gravitational radiation.  The star may eventually be
tidally disrupted, but on an orbit that is much more bound than in the
standard scenario.  This will lead to a remnant disc of the type analyzed
by \citet{CannizzoEtAl90}, for which the accretion rate decreases more
slowly than in the parabolic scenario: ${\dot M}\sim t^{-1.2}$ for
reasonable opacities.  If flare light curves are sampled sufficiently
these decays could in principle be distinguished from each other, which
would give us new insight into stellar dynamics and the prospects for
EMRIs.

Here we present numerical and analytical analyses of binary tidal
separation and subsequent tidal disruption of the remaining star.  We
note that there exist similar but not identical numerical studies. In
particular \cite{GouldQuillen2003} use a mass for the black hole of
$3.6\times 10^6\,M_{\odot}$ but show results only for the subset that
give captured stars with similar parameters to the observed stars
S2-0.  Their initial binary distributions are similar to ours,
although they do not examine binaries with initial semi-major axis
$<1$ AU and focus on higher masses.  \cite{GinsburgLoeb2006} address
a black hole mass of $4\times 10^6\,M_{\odot}$ and their binaries are
formed of two stars of masses $3\,M_{\odot}$. They present a few
sample orbits of captured stars similar to the S-stars, but do not
give a detailed distribution. \cite{PeretsGualandris2010} also focus
on $4\times 10^6\,M_{\odot}$ MBHs, and find as expected that the
captured stars tend to have high eccentricities $e>0.97$, but do not
give a periapsis distribution for the stars. \cite{MadiganEtAl2009}
present in their notable work direct-summation $N-$body simulations of small discs of stars
with semi-major axes of 0.026 and 0.26 pc with $4\,10^6\,M_{\odot}$
MBHs, which produced stars with high eccentricities that did not,
however, enter the region of greatest interest to us.  Hence we have
performed new numerical simulations to explore our scenario.

In \S~2 we discuss
tidal separations and present our three-body simulations of the process.  In
\S~3 we use these results as initial conditions and analyze the competition
between stellar dynamical processes (which can raise or lower the eccentricity)
and gravitational radiation (which shrinks and circularises the orbit) to
determine the mass ranges most likely to lead to moderate eccentricities at the
point of disruption.  In \S~4 we discuss the tidal process itself, and argue
that the small but nonzero residual eccentricities mean that for sufficiently
low-mass SMBHs the star will typically be disrupted rather than settling into a
phase of steady mass accretion onto the SMBH.  We present our conclusions in
\S~5.

\section{Binary tidal separation}
\label{sec_sep}

Tidal separation of binaries by SMBHs was first discussed by \cite{Hills88}. He
suggested that one member of the binary would be ejected with a velocity of $>
10^3\,\kms$, a ``hypervelocity star'' (HVS); several such objects have now been
observed (see \citealt{BrownEtAl09} for a discussion of their observed
properties).  The other member would settle into a fairly tightly bound orbit
around the SMBH; see \cite{MillerEtAl05} for a discussion in the context of
extreme mass ratio inspirals into an SMBH.

To simulate this process we assume a uniform distribution of pericentre
distances between 1 and 700 AU for the orbit of the binary around a
$10^7~M_\odot$ MBH.  The initial  orbit is
also assumed to be parabolic and to have its relative inclination uniformly
distributed over a sphere.  In total 228,000 numerical simulations were
conducted using a generalized three-body code described by
\cite{Zare74} and \cite{AarsethZare74}.  This numerical integrator is based on
Kustaanheimo-Stiefel regularisation of a two-body system, which is described in
\cite{KS65,Aarseth03}. The total energy and angular momentum of the system are
conserved to a high degree of accuracy and close encounters between bodies do
not induce unphysical velocities. 

\begin{figure}
\resizebox{\hsize}{!}{\includegraphics[scale=1,clip]{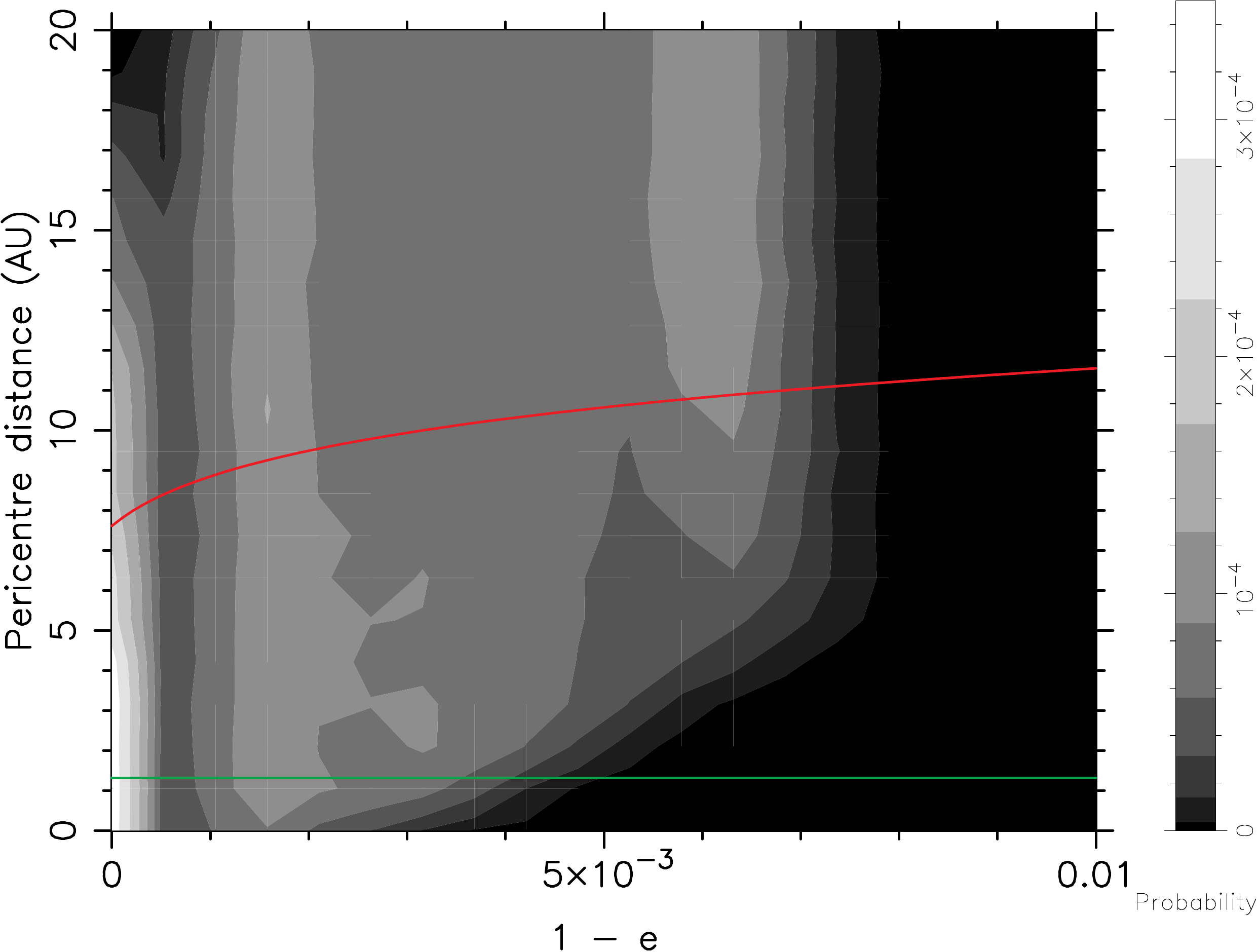}}
\caption{Distribution of the pericentre distance and eccentricity of the
captured companion at the tidal separation radius for an initial eccentricity
of $e_i=0.4$ and stellar mass of $1 \Msun$. Other eccentricities do not change
significantly the shape of the
distribution. The red line indicates the
maximum pericentre distance for which the tidal disruption happens within 
a Hubble time under the influence of gravitational radiation alone.
In the limit $e \rightarrow 1$, $r_p$ approaches the tidal disruption radius,
which we display as a green line, at at 1.3 AU, although this cannot be seen directly in the figure
because we are using a resolution of $\delta e = 10^{-4}$.
\label{fig.DistPeriapsisAndEcc}
}
\end{figure}

The resulting distribution for the pericentre distance and eccentricity of the
captured population, as well as the velocity distribution for the star
re-ejected into the stellar system, are shown in
Fig.(\ref{fig.DistPeriapsisAndEcc}) for an initial internal binary eccentricity
of $e_i=0.4$ and stellar mass of $1 \Msun$.  To produce this figure we chose
$10^7$ sets of parameters for fixed eccentricities and drew the semimajor axis
of the initial stellar binary from a log normal distribution between 0.05 and
10 AU. This is taken from observations of period distribution of binaries in
local field stars \citep{DM91}. The mean would be about 0.37 AU.

\begin{figure}
\resizebox{\hsize}{!}{\includegraphics[scale=1,angle=0,clip]{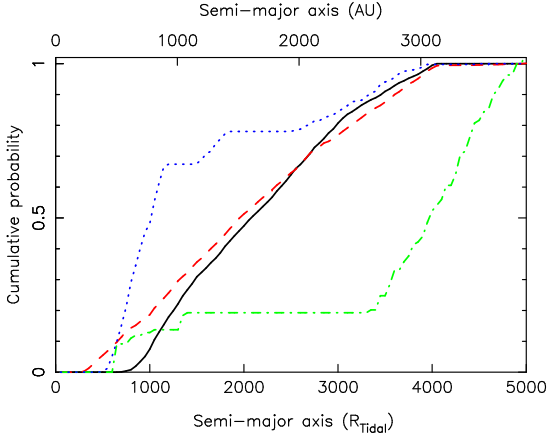}}
\caption{
Cumulative probability distribution of the semimajor axis of the captured star
after the tidal separation of the binary for a $1 \Msun$ star. The colours
denote the initial eccentricity of the binary before being disrupted by the
MBH, where black (solid line) is $e_i = 0.0$, red (dashed line) is $e_i = 0.4$, green (dot-dashed line) is $e_i = 0.7$ and
blue (dotted line) is $e_i = 0.9$. The probabilities of captures are different
in the different eccentricity cases, in particular the case $e_i=0.9$ is easier to capture
than the others.
\label{fig.CumulativeProb}
}
\end{figure}

In the figure we show the resulting probabilities, where we plot the probability of finding a
captured star with a particular pericentre and eccentricity bin given that a
binary is scattered to within 700 AU of the MBH.  The distribution of semimajor
axes for captured stars is shown in Fig.(\ref{fig.CumulativeProb}) for a $1
\Msun$ star that was taken from an initial stellar binary with eccentricity
$e_i = 0.0$, 0.4, 0.7 or 0.9.

We now discuss the evolution of the orbits of the stars after capture, under
the combined influence of two-body relaxation and gravitational radiation.

\section{Competition between stellar dynamics and gravitational radiation}

Suppose that a binary has been tidally separated by a close passage to a
supermassive black hole, but that the remaining object is outside the tidal
radius (i.e., it is not torn apart yet).  Gravitational radiation will
circularise the orbit as it shrinks, but dynamical processes can increase the
eccentricity.  Eventually, the star will move inside the tidal radius and (as
we argue in \S~4) will probably be tidally disrupted if the SMBH is
sufficiently low-mass.

In this section we discuss the dynamics subsequent to a tidal separation.  We
presume that the pericentre of the orbit of the remaining star is outside the
tidal radius, so that there is no immediate tidal disruption.  The star will
then be subjected to two-body interactions that can change the semimajor axis
and eccentricity of its orbit.  In principle resonant relaxation \citep{RT96}
could also play a role, in particular due to the high eccentricity the orbit
has, since the component of the torque is linearly proportional to eccentricity
\citep{GurkanHopman07}, but for the relevant tight orbits general relativistic
pericentre precession essentially eliminates this effect \citep{MerrittEtAl11}.
We will therefore focus exclusively on two-body interactions.  

The two-body energy relaxation time (during which the semimajor axis of the
orbit will be roughly doubled or halved) for a star of mass $m$ moving against
a background of density $\rho$ and velocity dispersion
$\sigma$ is \citep{Spitzer87}

\begin{equation}
t_{\rm en}\approx{0.3\over{\ln\Lambda}}{\sigma^3\over{
G^2\rho m}}\; .
\end{equation}
Here $\ln\Lambda\sim 10$ is the Coulomb logarithm.  
For our purposes, however, it is not the semimajor axis but
the pericentre distance that is important, because this is
what determines whether the star enters the tidal region.
It is therefore the angular momentum relaxation time that
is more relevant.  For a nearly circular orbit this time is
comparable to the energy relaxation time, but as we saw in
\S~\ref{sec_sep} the initial eccentricity is close to unity
in almost all cases.  The angular momentum of an orbit scales
as $\sqrt{a(1-e^2)}$, so an orbit with eccentricity $e$ has
an angular momentum a factor of $(1-e^2)^{1/2}$ less than a
circular orbit with the same semimajor axis.  Two-body
relaxation is a diffusive process, hence the expected change
in energy or angular momentum after time $t$ scales as
$t^{1/2}$.  As a result, the angular momentum relaxation time
is a factor of $[(1-e^2)^{1/2}]^2=1-e^2$ less than the energy
relaxation time:
\begin{equation}
t_{\rm am}={0.3\over{\ln\Lambda}}{\sigma^3\over{
G^2\rho m}}(1-e^2)\; .
\end{equation}
For $e\sim 1$ this is much shorter than the energy relaxation time,
hence we will assume that $a$ is fixed throughout.  We also note
that because angular momentum relaxation is a random walk process
the angular momentum could go up or down; if it goes up then nothing
interesting happens to the star, hence we will consider only the case
in which the angular momentum and hence the pericenter distance
decreases.

To be more quantitative, let us suppose that we have a galactic center with a
supermassive black hole of mass $M$ with a stellar mass density profile
$\rho(r)=\rho_0(r/r_{\rm infl})^{-\alpha}$ inside the radius of influence
$r_{\rm infl}\equiv 2GM/\sigma_0^2$, where $\sigma_0$ is the velocity
dispersion in the bulge of the galaxy.  The radius of influence is by
definition the radius inside of which the total stellar mass equals the black
hole mass, hence the normalization is $\rho_0={3-\alpha\over{4\pi}}
{M\over_{r_{\rm infl}^3}}$.  Suppose we make the simplifying approximation that
the velocity dispersion is $\sigma(r)=\sigma_0(r/r_{\rm infl})^{-1/2}$ (this
scaling is accurate for $r\ll r_{\rm infl}$ but not for $r\sim r_{\rm infl}$
because of the mass contribution from stars).  Let us assume in addition an
$M-\sigma_0$ relation of the form $M=10^8~M_\odot(\sigma_0/200~{\rm
km~s}^{-1})^4$ \citep{TremaineEtAl02}.  Then $r_{\rm infl}\approx 3
M_7^{1/2}$ pc, where $M=10^7M_7~M_\odot$, and

\begin{align}
\label{timescaleam}
t_{\rm am}& \approx 7\times 10^{11}~{\rm yr}(3-\alpha)^{-1}
M_7^{5/4}m_0^{-1}(r/r_{\rm infl})^{\alpha-3/2} \nonumber \\
          & (1-e^2)
\end{align}

\noindent
with $m_0\equiv m/M_\odot$.  
$t_{\rm am}$ is the timescale on which two-body processes can raise or
lower the pericentre distance significantly.  Competing against this is the
gravitational radiation timescale

\begin{equation}
t_{\rm GR}\approx 3\times 10^{15}~{\rm yr}~m_0^{-1}M_7^{-2}
\left(\frac{a}{1000 {\rm AU}}\right)^4 (1-e^2)^{7/2}\; .
\end{equation}

Over a time $t\approx t_{\rm GR}$, the orbit shrinks and circularises
significantly.  Setting the two timescales equal
to each other and noting that the pericentre distance is
$r_p=a(1-e)$ gives a critical pericentre distance of 

\begin{align}
r_{\rm p,\,crit}& \approx 16~{\rm AU}~(8\times 10^{-4})^{(2\alpha-3)/5} \nonumber \\
& (3-\alpha)^{2/5}M_7^{(8-\alpha)/5}
\left(\frac{a}{1000 {\rm AU}}\right)^{(2\alpha-6)/5}\; .
\end{align}

\noindent
Typical values for $r_{\rm p,\,crit}$ can be
read directly off of the simulations. For one that can
decay faster than a Hubble time it is $<10$ AU, and for one that can decay
faster than it would be disrupted by two-body relaxation it is more like 5 AU.
At a smaller pericentre distance than is given by this
expression, gravitational radiation dominates the evolution; conversely,
at a larger pericentre distance, two-body relaxation
dominates.

At the MBH masses $\sim 10^7~M_\odot$ that we consider, there may
or may not be time for the stars to relax dynamically, hence it is
not clear which value of $\alpha$ to take.  If strong mass segregation
occurs then $\alpha=2$ is likely 
\citep{AlexanderHopman09,PretoAmaroSeoane10,Amaro-SeoanePreto11},
but flatter slopes may also be relevant, particularly if there has
been scouring by a previous massive black hole merger and the system
has not yet readjusted.  For a selection of slopes we find

\begin{align}
r_{\rm p,crit}&\approx 14~{\rm AU}M_7^{13/10}(a/1000~{\rm AU})^{-3/5},
& \alpha=3/2\nonumber \\
r_{\rm p,crit}&\approx 7~{\rm AU}M_7^{5/4}(a/1000~{\rm AU})^{-1/2},
& \alpha=7/4\nonumber \\
r_{\rm p,crit}&\approx 4~{\rm AU}M_7^{6/5}(a/1000~{\rm AU})^{-2/5},
& \alpha=2\; .
\end{align}

We will simplify by assuming that gravitational radiation is
unimportant until $r_p=r_{\rm p,\,crit}$, at which point it takes
over completely with no further influence from two-body effects.
If this is true, then the next question is whether $r_{\rm p,\,crit}$
is greater than the tidal radius.  If we focus on main sequence
stars of mass $m\lesssim M_\odot$, then over a wide range of masses
their radii are reasonably fit by $R_{\star}\approx
0.85R_\odot(m/M_\odot)^{2/3}$ \citep{DemircanKahraman91} 
and the tidal radius is

\begin{equation}
r_T\approx R_{\star}\left(\frac{3M}{m}\right)^{1/3}\approx 
1.3~{\rm AU}M_7^{1/3}m_0^{1/3}\; .
\end{equation}

\noindent
Thus we see that stars in this mass range will typically
enter the gravitational radiation regime before they are
tidally disrupted.
Given that the critical pericenter is just a few times the tidal
radius, and that many aspects of this calculation are uncertain,
it is quite possible that although tidal effects drop off very
sharply with distance they could have an impact on the orbit 
outside $r_T$.  An exploration of this possibility would require
careful hydrodynamic simulations, but for our purposes we will
assume that they are not dominant.

Assuming that this is the case, we can compute the eccentricity of the
orbit at the point that the pericentre distance equals $r_T$, when (as
we show in the next section) the star is likely to be tidally
disrupted instead of settling into a phase of steady accretion.  We
calculate the eccentricity by noting that to lowest (quadrupolar)
order, pure evolution via gravitational radiation conserves the
quantity

\begin{equation}
C=ae^{-12/19}(1-e^2)\left(1+{121\over 304}e^2\right)^{-870/2299}
\end{equation}

\noindent
\citep{Peters64}.  We saw in \S~\ref{sec_sep} that the initial eccentricity
after tidal separation is nearly unity, so $1+e\approx 2$.  From our
assumptions we also know that $r_p=a(1-e)=r_{\rm p,\,crit}$.  Finally, if we
assume that at the tidal radius the eccentricity is $e_T\ll 1$, so that
$a_T\approx r_T$, we get

\begin{align}
a_Te_T^{-12/19} & \approx 1.8r_{\rm p,\,crit} \nonumber \\
e_T             & \approx 0.4 \left(\frac{r_{\rm p,\,crit}}{r_T}
                  \right)^{-19/12}\; .
\end{align}
For our three slopes the eccentricity at the tidal radius is thus

\begin{align}
e_T&\approx 0.01~M_7^{-551/360}m_0^{19/36}(a/1000~{\rm AU})^{19/20},
& \alpha=3/2\nonumber \\
e_T&\approx 0.03~M_7^{-209/144}m_0^{19/36}(a/1000~{\rm AU})^{19/24},
& \alpha=7/4\nonumber \\
e_T&\approx 0.07~M_7^{-247/180}m_0^{19/36}(a/1000~{\rm AU})^{19/30},
& \alpha=2
\end{align}

We now explore the consequences of the star sinking inside the tidal
radius with this eccentricity, and argue that tidal disruption is
the most likely outcome if the SMBH has sufficiently low mass.  
We then demonstrate that tidal disruption
with a small eccentricity leads to a different light curve than the
more commonly considered tidal disruption of a star on a parabolic
orbit.

\section{Hydrodynamics near and inside the tidal radius}

Suppose that the star sinks gradually under the influence of gravitational
radiation towards the tidal radius.  The tidal stresses increase as $\sim
(R_{\star}/r)^6$, where $R_{\star}$ is the stellar radius and $r$ is the
distance from the SMBH.  Therefore the star will be flexed and distorted, and
internal modes will be excited as it sinks \citep[for a recent discussion and
simple model of this complicated process, see][]{Ogilvie09}.  If the energy
from these modes could be dissipated then the orbit would undergo tidal
circularisation and might end up in a stable mass transfer state.  However, the
energy that must be dissipated is significantly larger than the binding energy
of the star.  To see this, note that at the tidal radius $r_T$, we have
$r_T=(3M/m)^{1/3}R_{\star}$.  The binding energy of the star is
$E_{\star}\approx Gm^2/R_{\star}$.  The binding energy of the orbit is $E_{\rm
orb}\approx GMm/r_T$.  Circularisation of an  orbit with eccentricity $e$ at
constant angular momentum releases an energy $e^2E_{\rm orb}$, so the ratio of
released energy to stellar binding energy is

\begin{equation}
e^2 \frac{E_{\rm orb}}{E_{\star}} \approx e^2 \left(\frac{M}{m}\right) 
\left(\frac{R_{\star}}{r_T}\right) \approx 3^{-1/3}e^2 \left(\frac{M}{m}\right)^{2/3}\; .
\end{equation}

If $M\sim 10^7m$ the ratio is therefore $\sim 3\times 10^4e^2$.  From the previous
section we found $e\sim 0.01-0.07$ for $M=10^7~M_\odot$, so the energy required to
circularise the orbit would be $\sim 3-150$ times the binding energy of the star.
If this energy could be released slowly
this would cause no problems (note for comparison that in its lifetime
the Sun will radiate a few hundred times its binding energy).
However, the thermal (Kelvin-Helmholtz) time for solar-type stars is a few tens
of millions of years, much longer than the inspiral time in our case and thus
the tidal stresses will build up more rapidly than their mode energy can be
radiated.  

The competition is therefore between the time needed for
gravitational radiation to move the star into the tidal radius
(where mass transfer will ensue) and the time needed for
circularisation due to tidal dissipation to deposit a stellar
binding energy into the star and thus, presumably, to tidally
disrupt the star.  Note that \cite{AlexanderMorris03} discussed how
tidal energy could produce ``squeezars" with a different appearance
from normal stars, without destroying the stars if the pericentre
distance is sufficiently large.  Here we are interested in the
conditions for tidal destruction.

To evaluate this we adapt the expressions from \citet{LeconteEtAl10}
for the energy deposition rate of tidal dissipation 
in a planet due to its eccentric orbit around a star.  They find
\begin{equation}
{\dot E}_{\rm tides}=2K_p\biggl|N_a(e)-{N^2(e)\over{\Omega(e)}}\biggr|
\end{equation}
where 
\begin{equation}
N(e)={1+{15\over 2}e^2+{45\over 8}e^4+{5\over{16}}e^6\over{(1-e^2)^6}}\; ,
\end{equation}
\begin{equation}
N_a(e)={1+{31\over 2}e^2+{255\over 8}e^4+{185\over{16}}e^6+
{25\over{64}}e^8\over{(1-e^2)^{15/2}}}\; ,
\end{equation}
\begin{equation}
\Omega(e)={1+{3\over 2}e^2+{1\over 8}e^4\over{(1-e^2)^5}}\; ,
\end{equation}
and
\begin{equation}
K_p\approx {9\over 4}Q^{-1}\left(Gm^2\over R_*\right)
\left(M\over m\right)^2\left(R_*\over a\right)^6\left(GM\over a^3\right)^{1/2}
\end{equation}

In the last equation $Q$ is the quality factor of the star, a standard 
parameterisation of the rate of tidal effects on to the star.  The magnitude of
$Q$ is notoriously uncertain; values of $Q=10^{5-6}$ are commonly
used (see, e.g., \citealt{MillerEtAl09} for a recent example).
If we use the expression $a_T=(3M/m)^{1/3}R_*$ for the tidal
radius, this last expression reduces to
\begin{equation}
K_p\approx {1\over 4}Q^{-1}\left(Gm^2\over R_*\right)
\left(a\over a_T\right)^{-6}\left(GM\over a^3\right)^{1/2}\; .
\end{equation}
In the limit $e\ll 1$ we find $N_a(e)\approx 1+23e^2$,
$N^2(e)\approx 1+27e^2$, and $\Omega(e)\approx 1+{15\over 2}e^2$.
Thus
\begin{equation}
{\dot E}_{\rm tides}\approx {7\over 4}Q^{-1}e^2
\left(Gm^2\over R_*\right)\left(a\over a_T\right)^{-6}
\left(GM\over a^3\right)^{1/2}\; .
\end{equation}
Thus the time needed to circularise the available energy
$\sim e^2(M/3m)^{2/3}(Gm^2/R_*)$ at $a=a_T$ is
\begin{equation}
\begin{array}{rl}
T_{\rm circ,tide}&\approx e^2(M/3m)^{1/3}(Gm^2/R_*)\\
&\ \times [(7/4)Q^{-1}e^2(Gm^2/R_*)(GM/a^3)^{1/2}]^{-1}\\
&=3\times 10^7~{\rm s}~QM_7^{2/3}m_0^{-1/6}\\
\end{array}
\end{equation}
where in the second line we have substituted $R_*=0.85R_\odot m_0^{2/3}$
and $a_T=(3M/m)R_*$.  The circularisation time from gravitational
radiation alone, at $e\ll 1$, is
\begin{equation}
\begin{array}{rl}
T_{\rm circ,GW}&\approx (15/304)c^5a^4/(G^3\mu M^2)\\
&\approx 6\times 10^{11}~{\rm s}~M_7^{-2/3}m_0^{1/3}\\
\end{array}
\end{equation}
\citep{Peters64}, where in the last line we again substituted
in $a=a_T$.  Thus $T_{\rm circ,GW}/T_{\rm circ,tide}\approx
2\times 10^4 Q^{-1}M_7^{-4/3}m_0^{1/2}$, which for $Q\sim 10^{5-6}$
is typically less than unity, hence only a fraction 
$T_{\rm circ,GW}/T_{\rm circ,tide}$ of the circularisation energy
will go into tidal heating.  Note, however, that for lower masses
the eccentricity at the tidal radius is larger (scaling roughly
as $M^{-3/2}$ for our three power laws) and that the ratio of
circularisation energy to the internal binding energy scales
as $e^2$, meaning that the total energy dissipated tidally
scales as $\sim M^{-4}$, approximately.  Thus even for $Q=10^6$,
several times the stellar binding energy will be dissipated
for $M<3\times 10^6~M_\odot$.

If instead the SMBH mass is large, so that 
gravitational wave circularisation dominates over tidal
circularisation, we expect that the star will settle into a period of steady
mass transfer.  The rate would be such that it balances the inward movement due
to gravitational radiation, i.e., the characteristic time would be of order the
gravitational radiation time.  For our typical values, this is roughly $10^5$
years, implying a rate of $\sim 10^{-5}~M_\odot$ per year.  Even if the
luminosity is produced with an efficiency of 10\%, this would produce a
luminosity of only $\sim 10^{41}$~erg~s$^{-1}$, weak enough and steady enough
that it would not be distinguishable from a standard low-luminosity AGN.  We
therefore focus on the possibility that the star is tidally disrupted and that
its debris is subsequently accreted by the SMBH.

If a star is disrupted from a low-eccentricity orbit the evolution of its tidal
debris proceeds differently than if it is disrupted from a parabolic orbit.  To
see this, note that in the original argument of \citet{Rees88} it was
demonstrated that the spread in the binding energy of the debris is comparable
to the range in orbital binding energy from one side of the star to the other.
If $m/M\sim 10^{-7}$, therefore, the fractional spread is $\sim (m/M)^{1/3}\sim
10^{-2}$.  As a result, if an orbit with a pericentre $r\sim r_T$ has an
eccentricity $e \gtrsim 0.99$, the debris semi-uniformly samples binding
energies from zero to the binding energy of the original stellar center of
mass.  The assumption of exactly uniform sampling (equal mass for equal range
in binding energy) leads to a mass accretion rate that scales with time $t$ as
$t^{-5/3}$; this law is more generally obtained at late times even for more
realistic assumptions about stellar structure \cite[e.g.][]{LodatoEtAl09}.  In
contrast, if the spread in debris energies is much less than the average
binding energy (corresponding to $e\ll 0.99$ in our example), then to lowest
order the debris moves in a thin stream that intersects itself and settles
within a few orbits into a remnant disc.  

Such discs were studied by \cite{CannizzoEtAl90}, who found that for plausible
opacities the accretion rate would decay more gradually, e.g., ${\dot M}\propto
t^{-1.2}$ for Thomson scattering.  Moreover, because the debris would all be
bound to the SMBH (unlike for the parabolic case, where roughly half the
stellar mass escapes to infinity), the accretion rate could be quite
substantial for comparatively low-mass SMBHs.  For Thomson scattering, the
expressions from \cite{CannizzoEtAl90} lead to

\begin{equation}
{\dot M}=2\times 10^{23}~{\rm g~s}^{-1}\left(\frac{\alpha}{0.1}\right)^{4/3}
{\bar\rho}^{7/9}M_7^{-10/9} \left(\frac{\Delta M}{M_\odot}\right)^{5/3}\; .
\end{equation}

Here ${\bar\rho}$ is the average density of the star in units of
g~cm$^{-3}$, $\Delta M$ is the mass of the remnant disc (which will
initially be the mass of the star) and $\alpha$ is the
\citet{ShakuraSunyaev73} viscosity parameter. For $M_7\lesssim 1$ this
therefore has the possibility of shining at luminosities that are a
significant fraction of the Eddington luminosity $L_E=1.3\times
10^{45}M_7~{\rm erg~s}^{-1}$ assuming an efficiency $L/{\dot M}c^2=0.1$.

As pointed out to us by E.~S. Phinney (2010, personal communication), depending
on the very uncertain details of how tidal energy is deposited, is it possible
that there will be a gravitational wave signature that attends the
electromagnetic signature of disruption.  In particular, it is not well
established whether the tidal energy is deposited uniformly in the volume of
the star or primarily where most of the matter is (both of which would lead to
full disruption) or primarily in the envelope.  If the last occurs, then the
envelope would be stripped and lead to significant accretion with the
characteristic decay discussed above, but the dense core would survive and
could spiral in further.  This would lead to a coincident gravitational wave
signal that could be detected with the proposed {\it LISA} if the source is
close enough \citep{Freitag03}.

\section{Conclusions}

In his work, \cite{Hopman09} estimates that for a galactic nucleus
such as ours, the tidal separation rate of binaries which start far
away from the MBH is $\Gamma^{\rm GC}_{\rm tid\,sep} \sim
7\times10^{-7}(f_{b}|^{\rm GC}/0.05)\,{\rm yr}^{-1}$, where ``GC''
stands for Galactic Center and $f_b$ is the fraction of stars in
binaries. Fig.(6) of \cite{Hopman09} shows that the rate increases
when we go to higher energies, because the loss-cone is depleted,
allowing more binaries to ``survive'' in their way to the GC. 
\cite{YuTremaine03} estimate that the number is enhanced by an order
of magnitude by binaries {\it not} bound to the MBH.  More
remarkably, the event rates can be at least temporarily enhanced by
{\em many orders of magnitude} if one considers the role of massive
perturbers, such as giant molecular clouds or intermediate-mass black
holes, which can accelerate relaxation by orders of magnitude as
compared to two-body stellar relaxation \citep{PeretsEtAl07}. 
Another important potential boosting effect is the possibility that
the potential is triaxial and not spherically symmetric
\citep{PM02,PoonMerritt04,MerrittPoon04}.  Taking these effects into
account, we assume $\Gamma^{\rm GC}_{\rm tid\,sep} \sim
10^{-5}(f_{b}|^{\rm GC}/0.05)\,{\rm yr}^{-1}$.  The fraction of main
sequence stars that will eventually spiral into the SMBH after tidal
separation is at least a few percent, so a plausible estimate of the
total event rate for tidal disruptions of a single star originated by
a separated binary in a Hubble time is $\Gamma^{\rm GC} \sim
10^{-7}(f_{b}|^{\rm GC}/0.05)\,{\rm yr}^{-1}$, and it could be
higher.  
This rate is probably a
subset of the rate at which single stars are likely to encounter
SMBHs on parabolic orbits \citep[see][for a
discussion of such extreme mass ratio
inspirals]{Amaro-SeoaneEtAl07}.
It is therefore possible that events
with the $L\propto t^{-1.2}$ decay characteristic of low-eccentricity
disruption may have rates smaller or similar to events with the $L\propto
t^{-5/3}$ decay that is expected to be signatures of disruption of
single stars in galactic nuclei and that is consistent with the
initial decay of the recent Swift event Sw 1644+57 \citep{BloomEtAl11}.

\section*{Acknowledgments}

The authors thank the referee for several useful suggestions.  PAS thanks Maite
Miranda Cascales for the sudden and unexpected two hours break in a small
Berlin bar on the way home, which led to the first calculations of the paper
accompanied by a little bottle of weinbrand.  He is indebted to Jos{\'e}
Mar{\'\i}a Ib{\'a}{\~n}ez and his group for the invitation to visit the
Departament d'Astronomia i Astrof{\'\i}sica de la Universitat de Val{\`e}ncia,
as well as to Jorge Cuadra for his visit at the Universidad Cat{\'o}lica de
Chile, where part of this work was done.  MCM thanks Sterl Phinney for valuable
discussions.  MCM was supported in part by NASA grants NNX08AH29G and
NNX12AG29G.  PAS and MCM
are indebted to the Aspen Center for Physics and the organizers of the summer
2011 meeting. GFK thanks the AEI for a visit in 2011.

\label{lastpage}

\end{document}